\documentstyle[aps,twocolumn,prl,epsf]{revtex}
\begin{document}

\twocolumn[\hsize\textwidth\columnwidth\hsize\csname
@twocolumnfalse\endcsname

\title
{Orbital ordering, Jahn-Teller distortion, and anomalous
 X-Ray scattering in Manganates}
\author {I.S. Elfimov$^1$, V.I. Anisimov$^1$ and G.A. Sawatzky$^2$}
\address {
$^1$Institute of Metal Physics, Russian Academy of Sciences, 620219
Ekaterinburg GSP-170, Russia\\
$^2$Laboratory of Applied and Solid State Physics,
Materials~Science~Center, University~of~Groningen,
Nijenborgh~4, 9747~AG~Groningen, The~Netherlands}

\maketitle

\begin{abstract}
We demonstrate with LSDA+U calculations that x-ray scattering at the
K edge of Mn is sensitive to orbital ordering in one energy range and
Jahn-Teller distortion in another. Contrary to what is suggested by
atomic or cluster models used  to date we show that band structure
effects rather than local Coulomb interactions dominate the
polarization dependence of the K edge scattering and therefore it is
sensitive to nearest neighbour bond length distortions and next
nearest neighbour orbital occupation. Based on this we propose a new
mechanism for K edge x-ray scattering in the manganates which we
suggest is also applicable to transition metal compounds in general.
\end{abstract}


%61.10.Dp Theories of diffraction and scattering
%71.27.+a Strongly correlated electron systems; heavy fermions
%71.20.-b - Electron DOS and band structure of crystalline solids

]

A very popular topic in strongly correlated 3$d$ transition metal systems
has to do with the influence of orbital degeneracy and orbital ordering
on the magnetic and magneto-electronic properties. It is well established
in the early works of Khomskii and Kugel\cite{KKh} that the superexchange
interactions between neighboring transition metal ions is strongly dependent
on the spatial orientation of occupied $d$ orbitals leading to even sign
changes in this interaction for different orbital occupations. Very much
studied recent examples are the so called colossal magnetoresistance
materials\cite{gmr} containing trivalent Mn ions with four 3$d$ electrons
which in a cubic crystal field and the usual strong atomic Hunds rule
coupling results in an electronic configuration with three  electrons with
parallel spins in the threefold degenerate t$_{2g}$ level and the fourth
electron in a twofold orbitally degenerate e$_g$ level. The remaining
twofold orbital degeneracy is in accordance with the Jahn-Teller theorem
lifted either by local lattice  distortions and or by ordering the e$_g$
level occupation on neighboring ions and thereby strongly affecting the
magnetic structure which in its turn stabilizes a particular orbital
ordering. Also the electrical transport properties are strongly affected
because of the so called double exchange like mechanisms in which the
electron band widths are strongly influenced by the spin structure and also
of course by the relative orientation of occupied orbitals on neighboring
ions\cite{Khomskii97}. Similar situations can and do occur in Fe and Co
oxides which also are recently a subject of intense investigation. A
somewhat different influence of orbital ordering has been invoked to
explain the very strange magnetic behavior of various V compounds like
LiVO$_2$ with its transition from a paramagnetic state to a non magnetic
state at 500K\cite{Pen97} and YVO$_3$ with its magnetization reversal
transitions at 90K and 77K\cite{Goodenough95,Ren98}. So the problem of orbital
ordering is quickly becoming a central issue in a broad range of materials.
A problem is to find an accurate and direct experimental method to determine
the nature of the orbital ordering and also study the changes which occur
at magnetic and crystallographic transitions.

Orbital ordering manifests itself in the site dependent orientation of the
quadrupole moment resulting from the spatial distribution of the  outermost
valence $d$ electrons. Unfortunately x-ray scattering under normal
conditions is primarily controlled by the core electrons of atoms and the
sensitivity to the valence electron distribution is usually very low.
However as has been demonstrated recently for the manganates
\cite{Murakami98a,Murakami98b} the use of x-ray energies corresponding to
K absorption edges can greatly enhance the sensitivity of x-ray scattering
cross sections to the valence electron distribution making a direct
observation of orbital ordering possible. Since we really want to study the
3$d$ electron distribution the use of an absorption edge corresponding to a
transition directly into the empty $d$ states would have the most
sensitivity but for the 3$d$ transition metals these involve 2$p$ or 3$p$
like core levels which are relatively shallow and therefore involve long
wavelengths restricting the information obtainable to systems with large
lattice parameters. In the above mentioned K edge experiments the direct
transitions to the $d$ states are not dipole allowed and therefore involve
weak quadrupole transitions. So where does the sensitivity come from?
Several models have been proposed one indeed involving the quadrupole
transitions and the other involving transition to the empty 4$p$ band
states which will be influenced by the $d$ electron occupation because of
the $d$-$p$ Coulomb interactions. Ishihara {\it et al} recently used a
nearest-neighbors MeO$_6$ octahedron to demonstrate the sensitivity of
the Mn 4$p$ states to the $d$ electron orbital occupation establishing a
basis for this effect\cite{Ishihara98}. However in this model the 4$p$
states are atomic like levels whereas in the real solid the 4$p$ states
form very broad bands which would tend to wash out the influence of the
local $d$-$p$ Coulomb interactions at first glance. For the analysis of
the data it is important to have a good understanding of the origin of the
effect in detail.

In this paper we present the results of a band structure study of the
effect of orbital ordering on the 4p density of states and especially the
local symmetry projected density of states. We find indeed that the 4$p$
bands are much broader than the $d$-$p$ Coulomb interactions so that this
is unlikely to be the dominant effect. However because the extended 4$p$
states so strongly hybridize either directly or via the O states, with
the 3$d$ states of the neighboring atoms the local $p_x$,$p_y$, and $p_z$
projected density of states is extremely sensitive to the local distortion
of the oxygen octahedron and the $d$ orbital occupation of the neighboring
Mn atoms. So by using the polarization and energy dependence of the
scattering, as was done in the experiment one can probe the orbital
orientation of the occupied $d$ states on the neighboring Mn atoms at
energies corresponding to the empty $d$ bands and  the Jahn-Teller
distortion of the oxygen octahedron at energies corresponding to the 4$p$
bands. This mechanism is quite different from that previously reported
since it is not the core hole parent atoms own orbital orientation which we
claim is measured but rather that of the \underline{neighboring} atoms. In
addition we present calculation both with and without the local Jahn-Teller
distortions so that the effects of a lattice distortion and a pure orbital
ordering effect can be separated.

In order to address those problems we performed LSDA+U calculations
on the prototypical orbital ordered compound LaMnO$_3$ and especially
studied the Mn-4$p$-DOS corresponding to p$_x$ and p$_y$ states.
It is necessary to realize that in standard LSDA orbital polarization can be
obtained only if MeO$_6$ octahedrons are distorted due to the Jahn-Teller
effect. However if one takes into account on site $d$-$d$ Coulomb
interaction, the resulting ground state is an orbitally polarized insulator
even for the crystal structure without Jahn-Teller distortion
\cite{Anisimov97}.

LaMnO$_3$ has a orthorhombic $Pbnm$ crystal structure which can be
considered as a cubic perovskite with two types of distortions: the first
one is a tilting of the MnO$_6$ octahedra, so that Mn-O-Mn angles become less
then 180$^{\circ}$, and the second one is a Jahn-Teller distortion (shortening
of the two Mn-O bond and elongation of a third one). The later is usually
considered responsible for the insulator ground state in LaMnO$_3$. The
configuration of a Mn$^{+3}$ ion in this compound is a high-spin $d^4$ state
which is represented by $3t_{2g\uparrow}$ and $1e_{g\uparrow}$ electrons.
Because of weak hybridization of $t_{2g}$ orbitals with O2p these states can
be regarded as forming the localized spin 3/2. In contrast to that, $e_g$
orbitals, which hybridize much more strongly with O2p produce in the end
rather broad bands. The strong exchange interaction with $t_{2g\uparrow}$
subshell leads to the splitting of the $e_g$ band into unoccupied
$e_{g\downarrow}$ and half-occupied $e_{g\uparrow}$ subbands.

The electronic structure of undoped LaMnO$_3$ was calculated by the
LSDA+U method in LMTO calculation scheme\cite{lmto} (based on Stuttgart
TBLMTO-47 computer code) with the values of U and J equal 8 and 0.88 eV
for Mn 3$d$ electrons, respectively. For the Mn atoms, the 
\begin{figure}
\epsfxsize=6.9cm
\centerline{\epsffile{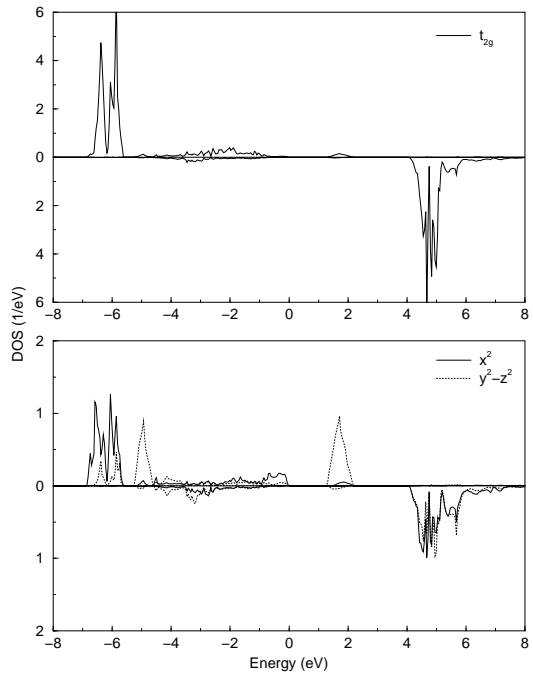}}
\vspace{2mm}
\caption{LSDA+U partial DOS of the Mn 3$d$ orbitals in LaMnO$_3$.
Upper half denotes majority spins and lower half the minority spin states
in each case. The zero of energy is at the Fermi energy.}
\label{Mn3d_DOS_in_La}
\end{figure}
\noindent
basis from 4$s$,
4$p$, and 3$d$ orbitals were used, while for La and O atoms it was 6$s$,
6$p$, 5$d$, 4$f$ and 2$s$, 2$p$, 3$d$ states, respectively.

\begin{figure}
\epsfxsize=7cm
\centerline{\epsffile{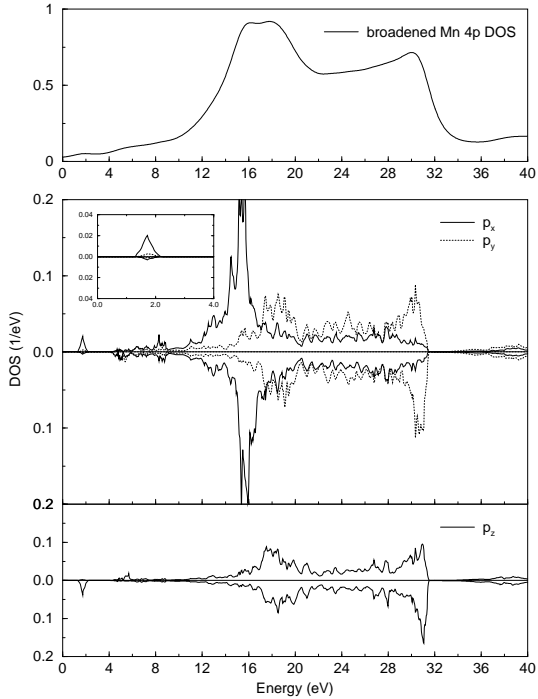}}
\vspace{2mm}
\caption{LSDA+U Mn 4$p$ partial DOS of the real structure. The upper
panel shows the 4$p$ total DOS broadened by a lorentzian of 2.4 eV
full width corresponding to the life time of the core hole state.
The middle and lower panels show the unbroadened p$_x$,p$_y$,
and p$_z$ spin projected density of states. Again upper panels are
majority spin states and lower panels minority spin states. The inset
in the center panel shows the 4$p$ partial density of states in the
Mn 3$d$ e$_g$ majority spin energy region.}
\label{Mn4p_DOS_in_La}
\end{figure}

As a first step we used the real crystal structure of the LaMnO$_3$
(with Jahn-Teller distortions). The result of the self-consistent
calculation was an orbital ordered antiferromagnetic insulator with a band
gap of 1.41 eV and the magnetic moment 3.84$\mu _B$ per Mn atom.
The $e_{g\uparrow}$-band is split by this gap into two subbands: the
occupied one with the predominantly $\phi_1$ character, and the empty one
with the $\phi_2$ character Fig.\ref{Mn3d_DOS_in_La}. Here, $\phi_1$ is a
$3x^2-r^2$ orbitals from the first type Mn atoms in the basal plane and
$3y^2-r^2$ of the second type (in the same plane) in a cubic coordinate
system which differ from the orthorhombic one by 45$^{\circ }$ rotation
around the $c$ axis. $\phi_2$ is  their $y^2-z^2$ and $z^2-x^2$ orbitals,
respectively. The 4$p$ partial DOS of the first Mn type is shown in
Fig.\ref{Mn4p_DOS_in_La} (the coordinate system is the same as in
Fig.\ref{Mn3d_DOS_in_La}). For the second Mn type the 4$p$ DOS has the
same character but the $p_x$ and $p_y$ states are interchanged.

\begin{figure}
\epsfxsize=6.85cm
\centerline{\epsffile{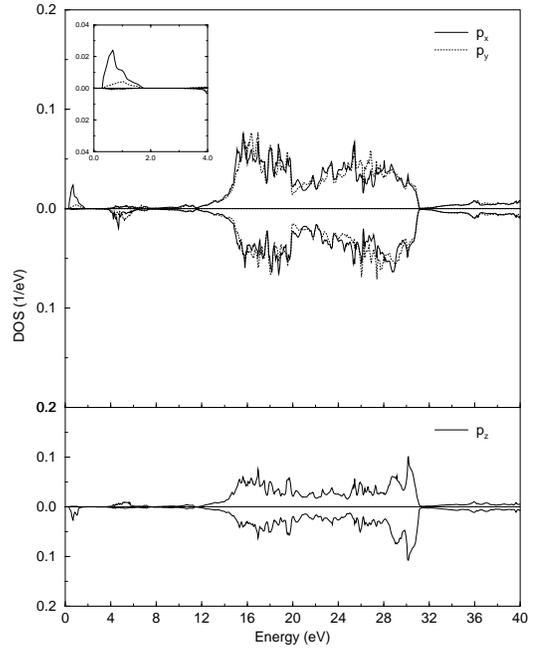}}
\vspace{2mm}
\caption{LSDA+U Mn 4$p$ partial DOS of LaMnO$_3$ in the crystal structure
without Jahn-Teller distortions.}
\label{Mn4p_DOS_in_Pr}
\end{figure}

\begin{figure}
\epsfxsize=6.9cm
\centerline{\epsffile{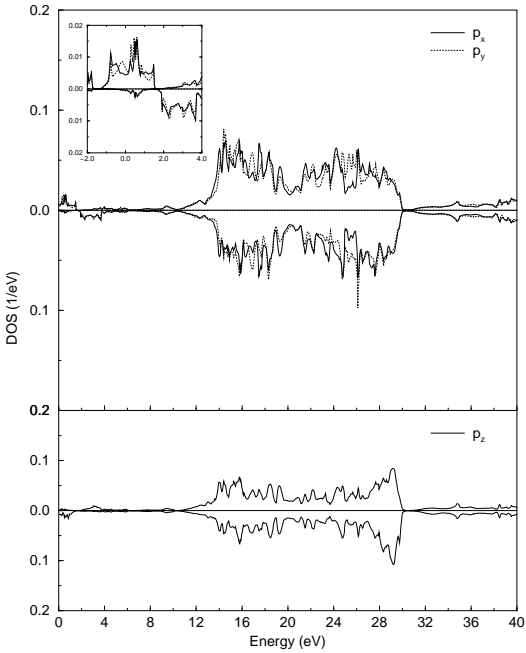}}
\vspace{2mm}
\caption{Mn 4$p$ partial DOS of LaMnO$_3$ in the crystal structure
without Jahn-Teller distortions and without orbital ordering as obtained
using LSDA.}
\label{Mn4p_lsdaDOS_in_Pr}
\end{figure}

Before we discuss these results in more detail we note that there are,
two interesting regions in the 4$p$ partial density of states and
therefore in the K edge x-ray absorption spectra: the first about 1.7 eV
above the Fermi energy corresponding to the Mn 3$d$ e$_g$ empty states
with some 4$p$ character mixed in and the second between 12 and 32 eV
corresponding to states of mainly 4$p$ character.
We note here the very strong difference in the p$_x$ and 
p$_y$ density of states which is the reason for the strong 
\noindent
anomalous
polarization dependent scattering at the K edge. In order to understand
the reason for why these two regions are interesting we performed the
calculation of the LaMnO$_3$ electronic structure in the crystal structure
of the Pr$_{1/2}$Sr$_{1/2}$MnO$_3$ which does not have the Jahn-Teller
distortion \cite{Knizek92}. This allows to separate the contribution
of Jahn-Teller distortion in the polarization of 4$p$ states from the
influence of Mn 3$d$ shell. This calculation was made in the same way
as the previous one and the result was also an antiferromagnetic
insulator with the orbital ordering of Mn $e_g$ electrons but with a
smaller band gap (0.39 eV) and the magnetic moment 3.81$\mu _B$ per
Mn atom. The calculated $d$-orbital polarization for real and
Pr$_{1/2}$Sr$_{1/2}$MnO$_3$ crystal structures has the same character
\cite{Anisimov97} Fig.1, so, it is not shown here. The comparison of these
two calculations (Fig.\ref{Mn4p_DOS_in_La} and Fig.\ref{Mn4p_DOS_in_Pr})
allows us to conclude that the Mn-4$p$-orbital polarization in the second
energy region is caused mainly by the Jahn-Teller distortion of oxygen
octahedra and in the first one by pure orbital ordering of Mn 3$d$ electrons.
In order to confirm the later supposition we performed the LSDA
calculation of the LaMnO$_3$ in the crystal structure without Jahn-Teller
distortions and without the local Coulomb correction resulting in a metal 
without any orbital
\noindent
ordering and also almost without polarization of Mn 4$p$
states Fig.\ref{Mn4p_lsdaDOS_in_Pr}. The magnetic moment is 3.17$\mu _B$ per
Mn atom.

Recently, Muracami {\it et all.} have applied  anomalous x-ray scattering
techniques to the LaMnO$_3$. They observed a resonant-like and polarization
dependent character of the scattering intensity near the K edge of the Mn
ion in this compound. The comparison of calculated 4$p$ densities of states
Fig.\ref{Mn4p_DOS_in_La} and the measured fluorescence Fig.2 in
\cite{Murakami98b} shows that they are very similar: the broad total 4$p$
DOS has also two peaks (in the region 12-32eV) and the distance between them
is about 13eV. Next, according to this experiment, the LaMnO$_3$ has the same
type of the orbital ordering as was given by Goodenough and which we found in
our calculations too. Based on the splitting of the Mn 4$p$ states Muracami
{\it et all.} proposed a theoretical description of the scattering mechanism
but they did not specify the origin of this splitting. From the MeO$_6$
cluster calculations \cite{Ishihara98} it was shown that the main reason of
the Mn 4$p$ polarization consists in the Coulomb interaction between Mn 3$d$
and 4$p$ electrons. So, this should lead to the raising in the energy of the
$p_x$ orbital which is oriented along the direction of the occupied
3$x^2$-$r^2$ orbital (in the cubic coordinate system) but according to our
calculation the picture is completely different. First of all, we should
note again that the polarization of the Mn 4$p$ orbitals in the region of
12-32 eV above Fermi energy originates mainly from the Jahn-Teller distortion
of the oxygen octahedra which is indirectly connected with the orbital
ordering. Second, in contrast to the cluster calculation, the $p_y$ Mn
orbital which is perpendicular to the occupied 3$x^2$-$r^2$ orbital (in cubic
coordinate system
\noindent
and for one particular Mn atom) is almost absent in the
region about 1.7 eV above Fermi energy. It means that the hybridization of
this orbital with the neighboring Mn 3$d$ orbitals has a much stronger
influence on the band structure of 4$p$ states then the $d$-$p$ Coulomb
interaction on the same site.

Summarizing, the LSDA+U calculation for the undoped LaMnO$_3$
demonstrates that there are two main contributions to the polarization
dependence of the K edge scattering both of which involve the 4$p$
orbitals and band structure effects. The first is caused by the
hybridization of the Mn 4$p$ orbitals with the ordered Mn 3$d$ orbitals on
neighboring Mn ions either directly or via the intervening O orbitals.
This effect is therefore sensitive to the $d$ orbital occupation not on
the central Mn ion with the core hole but rather that of  the
\underline{neighboring} Mn ions and occurs at energies corresponding
to the empty Mn 3$d$ bands. We should note here that if the Mn atoms are
not in lattice sites with inversion symmetry the 4$p$ states can mix directly
with the 3$d$ states of the same atom. In this case orbital ordering of the
core hole parent atom will directly be measured. This would be the
case for example in the corundum structure like V$_2$O$_3$. The second
effect originates from the hybridization of the central Mn 4$p$ states
with states centered on the neighboring O ions and is therefore very
sensitive to the local Jahn-Teller distortion but at most weakly affected
by the $d$ orbital occupation of the Mn 3$d$ states. This effect is visible
at about 10 eV higher in energy corresponding to the threshold of the
predominantly 4$p$ band. We note that although we have here demonstrated
this for Mn the physics described is very general and will be applicable
to transition metal compounds in general although the details like energy
scales will depend on the system.

This investigation was supported by the Russian Foundation for
Fundamental Investigations (RFFI grants 96-15-96598 98-02-17275),
and by the Netherlands Organization for Fundamental Research on Matter
(FOM) with financial support by the Netherlands Organization for
the Advance of Pure Science (NWO). The Groningen group also
acknowledges the financial support of the EU via the TMR OXSEN network.

\end{document}